\documentclass{article}
\begin{document}
{\LARGE
\begin{center}
{\bf
Widths of tetraquarks with hidden charm}
\end{center}
}

\large

\begin{center}
\vskip3ex
S.M. Gerasyuta $ ^{1,2}$ and V.I. Kochkin $ ^1$

\vskip2ex
$ ^1$ Department of Theoretical Physics, St. Petersburg State University,
198904,

St. Petersburg, Russia

$ ^2$ Department of Physics, LTA, 194021, St. Petersburg, Russia
\end{center}

\vskip2ex

\noindent
E-mail: gerasyuta@SG6488.spb.edu

\vskip4ex
\begin{center}
{\bf Abstract}
\end{center}
\vskip4ex
{\large
The relativistic four-quark equations are found in the framework of
coupled-channel formalism. The dynamical mixing of the meson-meson states
with the four-quark states is considered. The four-quark amplitudes of
the tetraquarks with  hidden charm including $u$, $d$, $s$ and charmed
quarks are constructed. The poles of these amplitudes determine the masses
and widths of tetraquarks.


\vskip2ex

\vskip2ex
{\bf I. Introduction.}
\vskip2ex
The observation of the $X(3872)$ [1 -- 4], the first of the $XYZ$ particles
to be seen, brought forward the hope that a multiquark state has been
received. Maiani et al. advocate a tetraquark explanation for the $X(3872)$
[5, 6]. Belle Collaboration observed the $X(3940)$ in double-charmonium
production in the reaction $e^+ e^- \to J/ \psi +X$ [7]. The state,
designated as $X(4160)$, was reported by the Belle Collaboration in Ref. 8.

The discovery of the $X(3872)$ has triggered intensive theoretical studies
to understand the structure of this state [9 -- 14].

One of the main issues is to clarity the quantum numbers, especially, the
spin and the parity, which are key properties to understand the abnormally
small width. Several review papers, as for example [15, 16], discuss the
difficulty of interpreting these resonances as charmonium states.

In our papers [17 -- 19] relativistic generalization of the three-body
Faddeev equations was obtained in the form of dispersion relations in the
pair energy of two interacting particles. The mass spectra of $S$-wave
baryons including $u$, $d$, $s$, $c$ quarks were calculated by a method
based on isolating of the leading singularities in the amplitude.

We searched for the approximate solution of integral three-quark equations
by taking into account two-particle and triangle singularities, all the
weaker ones being neglected. If we considered such an approximation, which
corresponds to taking into account two-body and triangle singularities,
and defined all the smooth functions of the subenergy variables in the
middle point of the physical region of Dalitz-plot, then the problem was
reduced to the one of solving a system of simple algebraic equations.

In the recent papers [20 -- 22] the relativistic three-quark equations for
the excited baryons are found in the framework of the dispersion
relation technique. We have used the orbital-spin-flavor wavefunctions for
the construction of integral equations. We calculated the mass spectra of
$P$-wave single, double and triple charmed baryons using the input
four-fermion interaction with quantum numbers of the gluon [22].

In this paper the relativistic four-quark equations are found in the
framework of coupled-channel formalism. The dynamical mixing between the
meson-meson states and the four-quark states is considered [23 -- 25].
Taking the $X(3872)$ and $X(3940)$ as input we predict the masses and
widths of $S$-wave tetraquarks with hidden charm (Table I).

After this introduction, we discuss the four-quark amplitudes, which
contain the two charmed quarks (Sect. II). In Sect. III, we report our
numerical results (Tables I and II).

\vskip2ex
{\bf II. Four-Quark Amplitudes for the Tetraquarks with Hidden Charm.}
\vskip2ex
We derive the relativistic four-quark equations in the framework of the
dispersion relations technique.

The correct equations for the amplitude are obtained by taking into
account all possible subamplitudes. It corresponds to the division of
complete system into subsystems with the smaller number of particles.
Then one should represent a four-particle amplitude as a sum of six
subamplitudes

\begin{equation}
A=A_{12}+A_{13}+A_{14}+A_{23}+A_{24}+A_{34}\, . \end{equation}

This defines the division of the diagrams into groups according to the
certain pair interaction of particles. The total amplitude can be
represented graphically as a sum of diagrams.

We need to consider only one group of diagrams and the amplitude
corresponding to them, for example $A_{12}$. We shall consider the
derivation of the relativistic generalization of the Faddeev-Yakubovsky
approach [26, 27] for the tetraquark.

We shall construct the four-quark amplitude of $c \bar c u \bar u$
meson, in which the quark amplitudes with quantum numbers of $0^{-+}$
and $1^{--}$ mesons are included. The set of diagrams associated with
the amplitude $A_{12}$ can further be broken down into five groups
corresponding to subamplitudes: $A_1 (s,s_{12},s_{34})$,
$A_2 (s,s_{23},s_{14})$, $A_3 (s,s_{23},s_{123})$, $A_4 (s,s_{34},s_{234})$,
$A_5 (s,s_{12},s_{123})$, if we consider the tetraquark with
$J^{pc}=2^{++}$.

Here $s_{ik}$ is the two-particle subenergy squared, $s_{ijk}$ corresponds
to the energy squared of particles $i$, $j$, $k$ and $s$ is the system
total energy squared.

In order to represent the subamplitudes  $A_1 (s,s_{12},s_{34})$,
$A_2 (s,s_{23},s_{14})$, $A_3 (s,s_{23},s_{123})$,
$A_4 (s,s_{34},s_{234})$ and $A_5 (s,s_{12},s_{123})$ in the form of
dispersion relations it is necessary to define the amplitudes of
quark-antiquark interaction $a_n(s_{ik})$. The pair quarks amplitudes
$q \bar q\rightarrow q \bar q$ are calculated in the framework of the
dispersion $N/D$ method with the input four-fermion interaction [28 -- 30]
with quantum numbers of the gluon [31]. The regularization of the dispersion
integral for the $D$-function is carried out with the cutoff parameter
$\Lambda$.
The four-quark interaction is considered as an input [31]:

\begin{equation}
g_V \left(\bar q \lambda I_f \gamma_{\mu} q \right)^2 +
g^{(s)}_V \left(\bar q \lambda I_f \gamma_{\mu} q \right)
\left(\bar s \lambda \gamma_{\mu} s \right)+
g^{(ss)}_V \left(\bar s \lambda \gamma_{\mu} s \right)^2 \, . \end{equation}

\noindent
Here $I_f$ is the unity matrix in the flavor space $(u, d)$. $\lambda$ are
the color Gell-Mann matrices. Dimensional constants of the four-fermion
interaction $g_V$, $g^{(s)}_V$ and $g^{(ss)}_V$ are parameters of the
model. At $g_V =g^{(s)}_V =g^{(ss)}_V$ the flavor $SU(3)_f$ symmetry occurs.
The strange quark violates the flavor $SU(3)_f$ symmetry. In order to avoid
an additional violation parameters, we introduce the scale shift of the
dimensional parameters [31]:

\begin{equation}
g=\frac{m^2}{\pi^2}g_V =\frac{(m+m_s)^2}{4\pi^2}g_V^{(s)} =
\frac{m_s^2}{\pi^2}g_V^{(ss)} \, .\end{equation}

\begin{equation}
\Lambda=\frac{4\Lambda(ik)}
{(m_i+m_k)^2}. \end{equation}

\noindent
Here $m_i$ and $m_k$ are the quark masses in the intermediate state of
the quark loop. Dimensionless parameters $g$ and $\Lambda$ are supposed
to be constants which are independent of the quark interaction type. The
applicability of Eq. (2) is verified by the success of
De Rujula-Georgi-Glashow quark model [32], where only the short-range
part of Breit potential connected with the gluon exchange is
responsible for the mass splitting in hadron multiplets.

We use the results of our relativistic quark model [31] and write down
the pair quarks amplitude in the form:

\begin{equation}
a_n(s_{ik})=\frac{G^2_n(s_{ik})}
{1-B_n(s_{ik})} \, ,\end{equation}

\begin{equation}
B_n(s_{ik})=\int\limits_{(m_i+m_k)^2}^{\frac{(m_i+m_k)^2\Lambda}{4}}
\hskip2mm \frac{ds'_{ik}}{\pi}\frac{\rho_n(s'_{ik})G^2_n(s'_{ik})}
{s'_{ik}-s_{ik}} \, .\end{equation}

\noindent
Here $G_n(s_{ik})$ are the quark-antiquark vertex functions. The vertex
functions are determined by the contribution of the crossing channels.
The vertex functions satisfy the Fierz relations. All of these vertex
functions are generated from $g_V$, $g^{(s)}_V$ and $g^{(ss)}_V$.
$B_n(s_{ik})$, $\rho_n (s_{ik})$ are the Chew-Mandelstam functions with
cutoff $\Lambda$ and the phase spaces, respectively.

In the case in question, the interacting quarks do not produce a bound
state; therefore, the integration in Eqs. (7) -- (11) is carried out from
the threshold $(m_i+m_k)^2$ to the cutoff $\Lambda(i,k)$.
The coupled integral equation systems (the meson state $J^{pc}=2^{++}$
for the $c \bar c u \bar u$) can be described as:

\begin{eqnarray}
A_1(s,s_{12},s_{34})&=&\frac{\lambda_1 B_1(s_{12})  B_1(s_{34})}
{[1- B_1(s_{12})][1- B_1(s_{34})]}+
4\hat J_2(s_{12},s_{34},1,1) A_3(s,s'_{23},s'_{123}) \, ,\\
&&\nonumber\\
A_2(s,s_{23},s_{14})&=&\frac{\lambda_2 B_1(s_{23})  B_1(s_{14})}
{[1- B_1(s_{23})][1- B_1(s_{14})]}+
2\hat J_2(s_{23},s_{14},1,1) A_4(s,s'_{34},s'_{234})\nonumber\\
&&\nonumber\\
&+&2\hat J_2(s_{23},s_{14},1,1) A_5(s,s'_{12},s'_{123}) \, ,\\
&&\nonumber\\
A_3(s,s_{23},s_{123})&=&\frac{\lambda_3 B_2(s_{23})}{[1- B_2(s_{23})]}+
2\hat J_3(s_{23},2) A_1(s,s'_{12},s'_{34})\nonumber\\
&&\nonumber\\
&+&\hat J_1(s_{23},2) A_4(s,s'_{34},s'_{234})
+\hat J_1(s_{23},2) A_5(s,s'_{12},s'_{123}) \, ,\\
&&\nonumber\\
A_4(s,s_{34},s_{234})&=&\frac{\lambda_4 B_2(s_{34})}{[1- B_2(s_{34})]}+
2\hat J_3(s_{34},2) A_2(s,s'_{23},s'_{14})\nonumber\\
&&\nonumber\\
&+&2\hat J_1(s_{34},2) A_3(s,s'_{23},s'_{234}) \, ,\\
&&\nonumber\\
A_5(s,s_{12},s_{123})&=&\frac{\lambda_5 B_2(s_{12})}{[1- B_2(s_{12})]}+
2\hat J_3(s_{12},2) A_2(s,s'_{23},s'_{14})\nonumber\\
&&\nonumber\\
&+&2\hat J_1(s_{12},2) A_3(s,s'_{23},s'_{123}) \, ,
\end{eqnarray}

\noindent
where $\lambda_i$, $i=1, 2, 3, 4, 5$ are the current constants. They do not
affect the mass spectrum of tetraquarks. Here $n=1$ corresponds to a
$q \bar q$-pair with $J^{pc}=1^{--}$ in the $1_c$ color state, and $n=2$
defines the $q \bar q$-pairs corresponding to tetraquarks with quantum
numbers: $J^{pc}=0^{++}$, $1^{++}$, $2^{++}$. We introduce the integral
operators:

\begin{eqnarray}
\hat J_1(s_{12},l)&=&\frac{G_l(s_{12})}
{[1- B_l(s_{12})]} \int\limits_{(m_1+m_2)^2}^{\frac{(m_1+m_2)^2\Lambda}{4}}
\frac{ds'_{12}}{\pi}\frac{G_l(s'_{12})\rho_l(s'_{12})}
{s'_{12}-s_{12}} \int\limits_{-1}^{+1} \frac{dz_1}{2} \, ,\\
&&\nonumber\\
\hat J_2(s_{12},s_{34},l,p)&=&\frac{G_l(s_{12})G_p(s_{34})}
{[1- B_l(s_{12})][1- B_p(s_{34})]}
\int\limits_{(m_1+m_2)^2}^{\frac{(m_1+m_2)^2\Lambda}{4}}
\frac{ds'_{12}}{\pi}\frac{G_l(s'_{12})\rho_l(s'_{12})}
{s'_{12}-s_{12}}\nonumber\\
&&\nonumber\\
&\times&\int\limits_{(m_3+m_4)^2}^{\frac{(m_3+m_4)^2\Lambda}{4}}
\frac{ds'_{34}}{\pi}\frac{G_p(s'_{34})\rho_p(s'_{34})}
{s'_{34}-s_{34}}
\int\limits_{-1}^{+1} \frac{dz_3}{2} \int\limits_{-1}^{+1} \frac{dz_4}{2}
 \, ,\\
&&\nonumber\\
\hat J_3(s_{12},l)&=&\frac{G_l(s_{12},\tilde \Lambda)}
{[1- B_l(s_{12},\tilde \Lambda)]} \, \, \frac{1}{4\pi}
\int\limits_{(m_1+m_2)^2}^{\frac{(m_1+m_2)^2\tilde \Lambda}{4}}
\frac{ds'_{12}}{\pi}\frac{G_l(s'_{12},\tilde \Lambda)
\rho_l(s'_{12})}
{s'_{12}-s_{12}}\nonumber\\
&&\nonumber\\
&\times&\int\limits_{-1}^{+1}\frac{dz_1}{2}
\int\limits_{-1}^{+1} dz \int\limits_{z_2^-}^{z_2^+} dz_2
\frac{1}{\sqrt{1-z^2-z_1^2-z_2^2+2zz_1z_2}} \, ,
\end{eqnarray}

\noindent
here $l$, $p$ are equal to $1$ or $2$.

In Eqs. (12) and (14) $z_1$ is the cosine of the angle between the relative
momentum of particles 1 and 2 in the intermediate state and the momentum
of the particle 3 in the final state, taken in the c.m. of particles
1 and 2. In Eq. (14) $z$ is the cosine of the angle between the momenta
of particles 3 and 4 in the final state, taken in the c.m. of particles
1 and 2. $z_2$ is the cosine of the angle between the relative
momentum of particles 1 and 2 in the intermediate state and the momentum
of the particle 4 in the final state, is taken in the c.m. of particles
1 and 2. In Eq. (13) $z_3$ is the cosine of the angle between relative
momentum of particles 1 and 2 in the intermediate state and the relative
momentum of particles 3 and 4 in the intermediate state, taken in the c.m.
of particles 1 and 2. $z_4$ is the cosine of the angle between the relative
momentum of particles 3 and 4 in the intermediate state and that of the
momentum of the particle 1 in the intermediate state, taken in the c.m.
of particles 3 and 4.

We can pass from the integration over the cosines of the angles to the
integration over the subenergies.

The solutions of the system of equations are considered as:

\begin{equation}
\alpha_i(s)=F_i(s,\lambda_i)/D(s) \, ,\end{equation}

\noindent
where zeros of $D(s)$ determinants define the masses of bound states of
tetraquarks. $F_i(s,\lambda_i)$ determine the contributions of
subamplitudes to the tetraquark amplitude.

\vskip2ex
{\bf III. Calculation results.}
\vskip2ex
The model in consideration has only two parameters: the cutoff
$\Lambda=10.0$ and the gluon coupling constant $g=0.794$. These parameters
are determined by fixing the tetraquark masses for the $J^{pc}=1^{++}$
$X(3872)$ and $J^{pc}=2^{++}$ $X(3940)$ [25]. The quark masses of model
$m_{u,d}=385\, MeV$ and $m_s=510\, MeV$ coincide with the ordinary meson
ones in our model [18, 19]. In order to fix anyhow $m_c=1586\, MeV$, we
use the tetraquark mass for the $J^{pc}=2^{++}$ $X(3940)$. The masses
and widths of meson-meson states with the spin-parity $J^{pc}=0^{++}$,
$1^{++}$, $2^{++}$ are considered in Table I. In our paper we predicted
the tetraquark $s \bar s c \bar c$ with the spin-parity $J^{pc}=2^{++}$
and mass $M=4160\, MeV$. Belle Collaboration observed the similar state,
designated as $X(4160)$, with mass $M=(4156\pm 29)\, MeV$ and total width
of $\Gamma=139^{+113}_{-65}\, MeV$ [8]. We predicted also the scalar
tetraquark $u \bar u c \bar c$ with mass $M=3708\, MeV$, which was
considered by Gamermann et al. [33]. They used the extension of $SU(3)$
Lagrangian to $SU(4)$, with the appropriate $SU(4)$ flavor symmetry breaking
and described the structure of charmed resonances as dynamically generated
states. The plans to produce this particle at BEPC-II, open up the
possibility to see the $X(3700)$ resonance as a narrow peak around
$50\, MeV$ in the photon spectrum from the radiative decay $\psi(3770)$
into $X(3700)$ $\gamma$. The tetraquark $X(3700)$ with $J^{pc}=0^{++}$
can decay to $J/\psi \,\omega$, $\eta\eta_c$ and $\eta\eta'_c$.

The functions $F_i(s,\lambda_i)$ (Eq. (15)) allow us to obtain the overlap
factors $f$ for the tetraquarks. We calculated the overlap factors $f$
and the phase spaces $\rho$ for the reactions $X\to M_1 M_2$ (Table II).
We calculated the widths of the tetraquarks with hidden charm (Table I).
We considered the formula $\Gamma \sim f^2 \times \rho$ [34], there $\rho$
is the phase space.

The widths of the tetraquarks are fitted by the fixing width
$\Gamma_{2^{++}}=(39\pm 26) \, MeV$ [35] for the tetraquark $X(3940)$.
The results of calculations allow us to consider the three tetraquarks
as the narrow resonances. The calculated width $X(3872)$ tetraquark with
the mass $M=3872.2\, MeV$ [35] and the spin-parity $J^{pc}=1^{++}$ is about
$\Gamma_{1^{++}}=2.1\, MeV$. In our model we neglected with the mass
distinction of the $u$ and the $d$ quarks. We used the sum of the masses of
mesons $J/\psi$ and $\rho_0$ ($m_{J/\psi}+m_{\rho_0}=3871.8 \, MeV$) [35]
for the calculation of phase space of the decay
$\rho_0 \, J/\psi \to \pi^+ \pi^- \, J/\psi$. We believe that this mass
is very nearly equal to sum of the masses of $D^0$ and $D^{*0}$ mesons
($m_{D^0}+m_{D^{*0}}=3871.81\pm 0.36 \, MeV$) for the tetraquark with the
spin-parity $J^{pc}=1^{+-}$. This sum was recently been precisely measured
by the CLEO Collaboration [36]. The experimental value of total
width of the tetraquark $X(3872)$ with the spin-parity $J^{pc}=1^{++}$ is
less than $2.3\, MeV$ [35]. The total width of the tetraquark $X(4160)$
is predicted $\Gamma_{1^{++}_s}=25 \, MeV$. The tetraquark $X(3700)$ with
the spin-parity $J^{pc}=0^{++}$ decays to $\eta\eta_c$. The calculated
width of this state is equal to $\Gamma_{0^{++}}=143 \, MeV$. The
tetraquarks with the spin-parity $J^{pc}=0^{++}$, $1^{++}$
$s \bar s c \bar c$ have only the weak decays.

\vskip2.0ex
{\bf Acknowledgments.}
\vskip2.0ex

The work was carried with the support of the Russian Ministry of Education
(grant 2.1.1.68.26).

\newpage

\noindent
Table I. Masses and widths of tetraquark with hidden charm.

\vskip1.5ex
\noindent
Parameters of model: quark masses $m_{u,d}=385\, MeV$, $m_s=510\, MeV$,
$m_c=1586\, MeV$, cutoff parameter $\Lambda=10.0$, gluon coupling constant
$g=0.794$.

\vskip1.5ex

\noindent
\begin{tabular}{|c|c|c|c|}
\hline
Tetraquark & $J^{pc}$ & Mass ($MeV$) & Width ($MeV$) \\
\hline
$X(3700)$ & $0^{++}$ & $3708$ & $143$ \\
($c \bar c$)($u \bar u$) & & & \\
($u \bar c$)($c \bar u$) & & & \\
\hline
$X(3872)$ & $1^{++}$ & $\underline{3872.2}$ & $2.1$ \\
($c \bar c$)($u \bar u$) & & & \\
($u \bar c$)($c \bar u$) & & & \\
\hline
$X(3940)$ & $2^{++}$ & $\underline{3940}$ & $\underline{39}$ \\
($c \bar c$)($u \bar u$) & & & \\
($u \bar c$)($c \bar u$) & & & \\
\hline
$X(4160)$ & $2^{++}$ & $4160$ & $25$ \\
($s \bar s$)($c \bar c$) & & & \\
($s \bar c$)($c \bar s$) & & & \\
\hline
\end{tabular}

\vskip12ex

\noindent
Table II. Overlap factors $f$ and phase spaces $\rho$ of  tetraquark
with hidden charm.

\vskip1.5ex

\noindent
\begin{tabular}{|l|l|l|l|}
\hline
Tetraquark (channels) & $J^{pc}$ & $f$ & $\rho$ \\
\hline
$X(3700)$ $\eta\eta_c$ & $0^{++}$ & $0.421$ & $0.233$ \\
$X(3872)$ $J/\psi \, \rho$ & $1^{++}$ & $0.211$ & $0.0137$ \\
$X(3940)$ $J/\psi \, \rho$ & $2^{++}$ & $0.274$ & $0.149$ \\
$X(4160)$ $J/\psi \, \varphi$ & $2^{++}$ & $0.238$ & $0.124$ \\
\hline
\end{tabular}

\newpage
{\bf \Large References.}
\vskip5ex
\noindent
1. S.K. Choi et al. (Belle Collaboration), Phys. Rev. Lett. {\bf 91},
262001 (2003).

\noindent
2. D. Acosta et al. (CDF Collaboration), Phys. Rev. Lett. {\bf 93},
072001 (2004).

\noindent
3. V.M. Abazov et al. (D0 Collaboration), Phys. Rev. Lett. {\bf 93},
162002 (2004).

\noindent
4. B. Aubert et al. (BaBar Collaboration), Phys. Rev. D{\bf 71},
071103 (2005).

\noindent
5. L. Maiani, F. Piccinini, A.D. Polosa and V. Riequer,
Phys. Rev. D{\bf 71}, 014028

(2005).

\noindent
6. L. Maiani, A.D. Polosa and V. Riequer, Phys. Rev. Lett. {\bf 99},
182003 (2007).

\noindent
7. K. Abe et al. (Belle Collaboration), Phys. Rev. Lett. {\bf 98},
082001 (2007).

\noindent
8. I. Adachi et al. (Belle Collaboration), arXiv: 0708.3812 [hep-ex].

\noindent
9. F.E. Close and P.R. Page, Phys. Lett. B{\bf 628}, 215 (2005).

\noindent
10. T. Barnes, Int. J. Mod. Phys. A{\bf 21}, 5583 (2006).

\noindent
11. E.S. Swanson, Int. J. Mod. Phys. A{\bf 21}, 733 (2006).

\noindent
12. X. Liu, Z.G. Luo, Y.R. Liu and S.L. Zhu, arXiv: 0808.0073 [hep-ph].

\noindent
13. Y. Dong, A. Faessler, T. Gutsche and V.E. Lyubovitskij,
arXiv: 0802.3610 [hep-ph].

\noindent
14. C.E. Thomas and F.E. Close, arXiv: 0805.3653 [hep-ph].

\noindent
15. E.S. Swanson, Phys. Rept. {\bf 429}, 243 (2006).

\noindent
16. S. Godfrey and S.L. Olsen, arXiv: 0801.3867 [hep-ph].

\noindent
17. S.M. Gerasyuta, Yad. Phys. {\bf 55}, 3030 (1992).

\noindent
18. S.M. Gerasyuta, Z. Phys. C{\bf 60}, 683 (1993).

\noindent
19. S.M. Gerasyuta, Nuovo Cim. A{\bf 106}, 37 (1993).

\noindent
20. S.M. Gerasyuta and E.E. Matskevich, Yad. Fiz. {\bf 70}, 1995 (2007).

\noindent
21. S.M. Gerasyuta and E.E. Matskevich, Phys. Rev. D{\bf 76}, 116004 (2007).

\noindent
22. S.M. Gerasyuta and E.E. Matskevich, Int. J. Mod. Phys. E{\bf 17},
585 (2008).

\noindent
23. S.M. Gerasyuta and V.I. Kochkin, Z. Phys. C{\bf 74}, 325 (1997).

\noindent
24. S.M. Gerasyuta and V.I. Kochkin, Nuovo Cim. A{\bf 110}, 1313 (1997).

\noindent
25. S.M. Gerasyuta and V.I. Kochkin, arXiv: 0804.4567 [hep-ph].

\noindent
26. O.A. Yakubovsky, Sov. J. Nucl. Phys. {\bf 5}, 1312 (1967).

\noindent
27. S.P. Merkuriev and L.D. Faddeev, Quantum scattering theory for system
of few

particles (Nauka, Moscow 1985) p. 398.

\noindent
28. Y. Nambu and G. Jona-Lasinio, Phys. Rev. {\bf 122}, 365 (1961):
ibid. {\bf 124}, 246

(1961).

\noindent
29. T. Appelqvist and J.D. Bjorken, Phys. Rev. D{\bf 4}, 3726 (1971).

\noindent
30. C.C. Chiang, C.B. Chiu, E.C.G. Sudarshan and X. Tata,
Phys. Rev. D{\bf 25}, 1136

(1982).

\noindent
31. V.V. Anisovich, S.M. Gerasyuta, and A.V. Sarantsev,
Int. J. Mod. Phys. A{\bf 6}, 625

(1991).

\noindent
32. A.De Rujula, H.Georgi and S.L.Glashow, Phys. Rev. D{\bf 12}, 147 (1975).

\noindent
33. D. Gamermann, E. Oset and B.S. Zou, arXiv: 0805.0499 [hep-ph].

\noindent
34. J.J. Dudek and F.E. Close, Phys. Lett. B{\bf583}, 278, (2004).

\noindent
35. C. Amsler et al. (Particle Data Group),
Phys. Lett. B{\bf 667}, 1 (2008).

\noindent
36. C. Cawfield et al. (CLEO Collaboration), Phys. Rev. Lett. {\bf 98},
092002 (2007).

\end{document}